\documentclass[
 reprint,
 amsmath,amssymb,
 aps,prl,
 longbibliography,
 groupedaddress
]{revtex4-2}

\usepackage{mathptmx}           
\usepackage{graphicx}           
\usepackage{amssymb,amsmath}
\usepackage{booktabs}
\usepackage{multirow}
\usepackage{dcolumn}            
\usepackage{bm}                 
\usepackage{mathrsfs}
\usepackage{siunitx}
\usepackage{ORCIDinREVTeX}
\usepackage[hidelinks,
    colorlinks=true,
    linkcolor=blue,             
    citecolor=blue,             
    filecolor=blue,             
    urlcolor=blue]{hyperref}    

\renewcommand{\eqref}[1]{\hyperref[{#1}]{\textup{(\ref*{#1}})}}
\newcommand{\figref}[1]{\hyperref[{#1}]{\textup{Fig.~\ref*{#1}}}}
\newcommand{\secref}[1]{\hyperref[{#1}]{\textup{Sec.~\ref*{#1}}}}
\newcommand{\tabref}[1]{\hyperref[{#1}]{\textup{Table~\ref*{#1}}}}

\begin{document}

\preprint{APS/123-QED}

\title{Observation of higher-order exceptional points in pseudo-Hermitian radio-frequency circuits}

\author{Ke Yin}
\orcid{0000-0002-8534-216X}

\author{Xianglin Hao}
\orcid{0000-0001-7149-0113}

\author{Yuangen Huang}
\orcid{0000-0002-2486-2778}

\author{Jianlong Zou}
\orcid{0000-0003-1489-0828}

\author{Xikui Ma}
\orcid{0000-0003-4138-7443}

\author{Tianyu Dong}
\orcid{0000-0003-4816-0073}
\email[\vspace{-1.0em}Corresponding author: ]{tydong@mail.xjtu.edu.cn}
\affiliation{School of Electrical Engineering, Xi'an Jiaotong University, Xi'an 710049, China.}

\date{April 18, 2023}

\begin{abstract}
Exceptional points (EP) in non-Hermitian systems have been widely investigated due to their enhanced sensitivity in comparison to standard systems. In this letter, we report the observation of higher-order pseudo-Hermitian degeneracies in an electronic platform comprised of three inductively coupled gain-loss-loss LC resonators. Theoretical analysis demonstrates that the proposed system can realize third-order EP with asymmetric coupling between adjacent inductors and an arbitrary scaling factor between two loss resonators. When capacitive perturbation is introduced on the middle resonator, the perturbed eigenfrequencies follow a cube-root dependence on the perturbation parameter; in this case, the sensitivity is significantly greater than conventional wireless readout methods. Our work enriches the explorations of higher-order EP on electronic platforms and provides a new degree of design freedom for the non-Hermitian-EP-enhanced wireless sensing system.
\end{abstract}

\maketitle

An exceptional point (EP) is a singularity in the parameter space of the system where the eigenvalues and their corresponding eigenvectors coalesce \cite{miri2019exceptional}. In recent decades, the characteristics of EP in non-Hermitian systems, especially parity-time (PT)-symmetric systems, have been intensively studied in a variety of physical settings, including optical waveguides \cite{guo2009observation,doppler2016dynamically}, microresonators \cite{chang2014parity}, cavity magnon-polaritons \cite{zhang2017observation}, and coupled inductor-capacitor (LC) resonators \cite{schindler2012symmetric,zhou2021observation,zeng2019enhanced,dong2019sensitive,yin2022wireless}, to mention a few. The phase transition properties at EP are commonly associated with intriguing phenomena, such as asymmetric mode switching \cite{doppler2016dynamically}, single-mode lasing \cite{hodaei2014parity,feng2014single}, enhanced sensitivity to external perturbations \cite{hodaei2017enhanced,zeng2019enhanced}, \emph{etc}. Enhanced optical sensing systems based on second-order EP in non-Hermitian systems is first proposed for nanoparticle detection \cite{wiersig2014enhancing}. Later, higher-order EPs (HOEPs) with even higher sensitivities were proposed and experimentally validated \cite{hodaei2017enhanced}. By borrowing the concept from the pseudo-Hermitian condition for realizing non-Hermitian systems with real eigenvalues \cite{mostafazadeh2002pseudo,hao2023frequency}, some researchers propose that pseudo-Hermitian systems without PT symmetry can also exhibit HOEP \cite{xiong2021higher}. Until now, pseudo-Hermitian-EP properties have been mostly investigated in optical systems and have yet to be extended to radio-frequency (RF) electronic circuits. In our work, it is demonstrated that coupled LC resonators with pseudo-Hermiticity operated in HOEP can realize wireless sensing systems with enhanced sensitivity.

The idea of wirelessly monitoring physiological parameters through inductively interrogating LC mircosensors was first proposed by Collins to measure intraocular pressure \cite{collins1967miniature}. The conventional readout method is by inductively coupling the passive microsensor to a readout antenna, which is connected to a vector network analyzer (VNA) for reflection measurement. The measured parameter induces the resonant frequency change of the LC microsensor and successively introduces a dip frequency shift of the reflection spectra. During the last few decades, the microsensor design and manufacturing process itself has been greatly improved with the development of MEMS technology \cite{chen2008microfabricated,chen2010wireless,xue2012su8based}. However, fewer efforts have been made to improve the readout performance, \emph{i.e.}, resolution and sensitivity, of the sensing system, which has hindered the real-life application of the passive LC sensor. Recently, it has been proposed that a PT-symmetry-based wireless sensing system can improve the resolution of the readout spectrum \cite{chen2018generalized,yang2021ultrarobust,yin2022ultrahigh,zhou2020enhancing}. Furthermore, by utilizing the eigenvalue bifurcation properties of EP, the sensitivity of the response to the parameters to be measured can be enhanced \cite{zhou2021observation,zeng2019enhanced,dong2019sensitive,zeng2021ultra}. For instance, perturbed eigenvalues at second- and third-order EP with enhanced sensitivity have been observed in a RF PT-symmetric system \cite{zhou2021observation,zeng2019enhanced}. Along different lines, it is proposed that a second-order EP-locked reader with PT symmetry can interrogate an LC microsensor with enhanced sensitivity compared to the standard reader \cite{dong2019sensitive}, which has been extended to the HOEP-locked reader \cite{zeng2021ultra}. However, the realization of HOEP in PT-symmetric systems requires a critical system configuration condition. Taking third-order EP (EP3) as an example, typical PT-symmetric systems with EP3 consist of coupled gain-neutral-loss resonators, in which the neutral resonator has to be completely lossless and the coupling between adjacent resonators has to be equal. These critical conditions limit the design freedom of EP3-based sensors and increase the challenge of experimental verification. Fortunately, PT symmetry is only a sufficient but unnecessary condition for constructing HOEP \cite{xiong2021higher}. Instead, the more general pseudo-Hermitian wireless sensing systems can be designed to exhibit EP3, relaxing the condition of PT symmetry.

In this letter, we propose a pseudo-Hermitian wireless sensing system consisting of gain-loss-loss LC resonators. Theoretical analysis based on coupled mode theory shows that the proposed system can realize third-order EP under an asymmetric adjacent coupling condition, which has been experimentally verified. When capacitive perturbation is introduced on the relay resonator, the perturbed eigenfrequency follows a cube-root dependency on the perturbation parameter. The sensitivity at EP3 is much higher compared to the conventional wireless readout method. Our work demonstrates that EP-enhanced sensing can be utilized under arbitrary coupling and gain-loss parameters given that the pseudo-Hermitian condition is satisfied, offering a new degree of design freedom for the HOEP-based wireless sensing system.

\begin{figure}[htbp!]
    \centering
    \includegraphics[width = 3.4in]{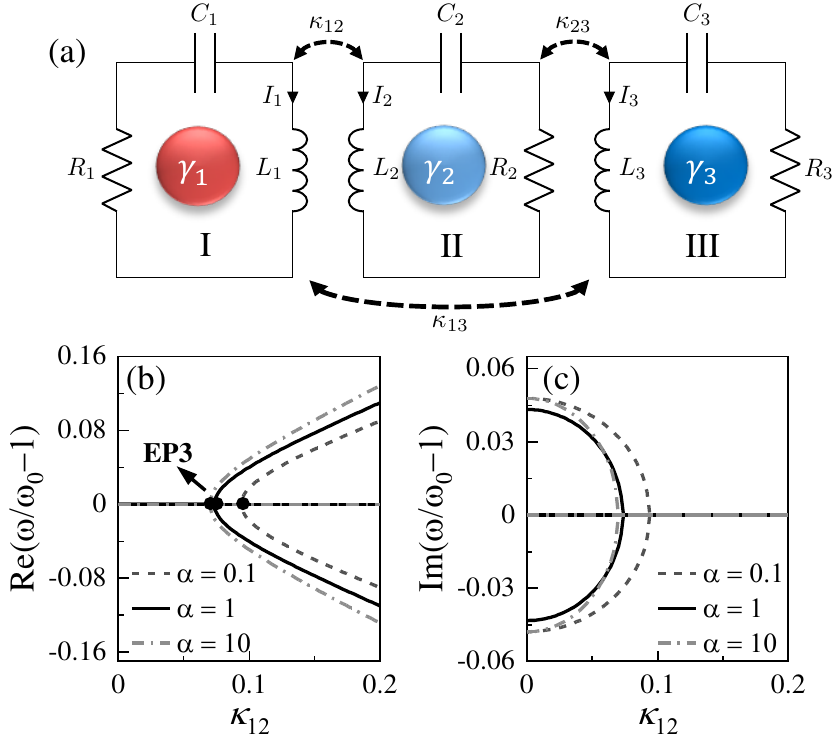}
    \caption{(a) Illustration and circuit schematic of the pseudo-Hermitian electronic trimer, which is composed of three planarly inductively coupled RLC resonators. (b) Real and (c) imaginary parts of eigenfrequency evolution as a function of $\kappa_{12}$ at different scaling factors $\alpha$ when $g = 0.1$, $\kappa_{13} = 0$ and $\kappa_{23} = (1+\alpha)^{-3/2}\kappa_{12}$.}
    \label{fig:fig01}
\end{figure}

The schematic of the proposed pseudo-Hermitian electronic trimer, consisting of three planarly inductively coupled RLC series resonators, is shown in \figref{fig:fig01}(a). For generality, we first assume all three resonators are lossy, with a loss parameter $\gamma_n$ ($n = 1,2,3$). By obtaining the conditions for HOEP, the gain and loss parameters can be determined. 

The system equation of the proposed system derived from coupled mode theory (CMT) reads $\text{i} \text{d}\bm{I}/\text{d}t = \hat{H}_0 \bm{I}$ with
\begin{equation}
    \hat{H}_0 = \frac{\omega_0}{2}
    \begin{pmatrix}
        2 - \text{i}\gamma_1   & \kappa_{12}            &   \kappa_{13}   \\
        \kappa_{12}            & 2 - \text{i}\gamma_2   &   \kappa_{23}  \\
        \kappa_{13}            & \kappa_{23}            &   2 - \text{i}\gamma_3
    \end{pmatrix},
\end{equation}
where $\bm{I} = (I_1, I_2, I_3)^\text{T}$ is the system variable; $\omega_0 = 1/\sqrt{L_n C_n}$ denotes the resonant frequency with $L_n = L$ and $C_n = C$ for $n=1, 2, 3$; $\gamma_n = R_n\sqrt{C/L}$ is the gain/loss parameter; $\kappa_n = M_n/L$ is the inductive coupling coefficient where $M_n$ is the mutual inductance; $\hat{H}_0$ is the effective Hamiltonian operator.

After a substitution $\lambda = \omega/\omega_0 - 1$, the real and imaginary parts of the characteristic equation $\det(\hat{H}_0 - \omega \mathbb{I}_3) = 0$, with $\mathbb{I}_3$ being a $3 \times 3$ identity matrix, respectively reads
\begin{subequations}
 \begin{align}
    \lambda^3 - \frac{1}{4}\sum_{m < n} (\gamma_m \gamma_n + \kappa_{mn}^2)\lambda - \frac{1}{4}\prod_{m < n}\kappa_{mn} &= 0,  \label{eq:char_eq_re} \\
    \sum_{m=1}^{3}\gamma_m \lambda^2 - \frac{1}{4}\prod_{m=1}^{3}\gamma_m \left(1 + \sum_{m < n}\frac{\kappa_{mn}^2}{\gamma_m \gamma_n} \right) &= 0, \label{eq:char_eq_im}
    \end{align}
\end{subequations}
where $m = 1, 2, 3$ and $n = 1, 2, 3$. Here $\omega_0$ is normalized to one for concision. With the characteristic equation, we can then determine the conditions for the system to be pseudo-Hermitian as well as the conditions for the emergence of third-order EP.

In order to guarantee pseudo-Hermiticity, the Hamiltonian should satisfy $\det(\hat{H}_0 - \omega \mathbb{I}_3) = \det(\hat{H}_0^* - \omega \mathbb{I}_3)$, \emph{i.e.} $\text{Im}[\det(\hat{H}_0 - \omega \mathbb{I}_3)] = 0$, which yields
\begin{subequations} 
 \begin{align}
    \sum_{m=1}^{3}\gamma_m & = 0, \label{eq:condition_a}  \\
    \prod_{m=1}^{3}\gamma_m \left(1 + \sum_{m < n}\frac{\kappa_{mn}^2}{\gamma_m \gamma_n} \right) & = 0. \label{eq:condition_b}
\end{align}
\end{subequations}
Here, the condition \eqref{eq:condition_a} implies that balanced total gain and loss of the system should be satisfied. As can been seen from \eqref{eq:char_eq_re}, the third-order EP $\lambda_\text{EP3}$ arises when 
\begin{subequations} \label{eq:condition_2}
 \begin{align}
    \sum_{m < n}(\gamma_m \gamma_n +  \kappa_{mn}^2) & = 0, \label{eq:condition_2a} \\
    \prod_{m < n}\kappa_{mn} & = 0. \label{eq:condition_2b}
 \end{align}
\end{subequations}
Taking into account the structure of the wireless sensing system and utilizing the gain introduced by the VNA, we consider the case where $\gamma_1 = -g$ and $\gamma_2 = \alpha \gamma_3$, \emph{i.e.}, the first resonator is gain while the remaining two resonators are lossy with loss parameters that differ by $\alpha$. From \eqref{eq:condition_a}, $\gamma_2$ and $\gamma_3$ can be derived as $\gamma_2 = g \alpha/(1 + \alpha)$ and $\gamma_3 = g/(1 + \alpha)$, respectively. Note that when $\alpha = 0$, the system is a PT-symmetric trimer with a neutral relay resonator. By solving \eqref{eq:condition_b} and \eqref{eq:condition_2}, it is possible to determine the coupling coefficients $\kappa_{12}, \kappa_{13}$ and $\kappa_{23}$ for a given gain parameter $g$. Note that the condition \eqref{eq:condition_2b} implies that the third-order EP exists only when one of the coupling coefficients equals to zero. Specifically, when $\kappa_{12} = 0$, the derived coupling coefficients are $\kappa_{13} = g \sqrt{(1+\alpha)/(1+2\alpha)}$ and $\kappa_{23} = g \alpha \sqrt{\alpha/(1+2\alpha)} /(1+\alpha)$, respectively; when $\kappa_{13} = 0$, we have $\kappa_{12} = g \sqrt{(1+\alpha)/(2+\alpha)}$ and $\kappa_{23} = g/[(1+\alpha)\sqrt{2+\alpha}]$; when $\kappa_{23} = 0$, there are no physical solutions. The coupling between the two lossy resonators should therefore be nonzero. For a PT-symmetric trimer where $\alpha = 0$, the EP3 condition becomes $\kappa_{12} = \kappa_{23} = g/\sqrt{2}$, which agrees with the previous work \cite{hodaei2017enhanced}.

Figures \ref{fig:fig01}(b) and \ref{fig:fig01}(c) plot the real and imaginary parts of the eigenfrequency evolution at various $\alpha$ with respect to $\kappa_{12}$, respectively, when $\kappa_{13} = 0$ and $\kappa_{23}=(1+\alpha)^{-3/2}\kappa_{12}$. Here, the gain parameter is $g = 0.1$. It can be seen that in the weak coupling regime when $\kappa_{12}<\kappa_\text{EP3}$, the system has one real eigenvalue and a pair of complex conjugate eigenvalues, whereas in the strong coupling regime when $\kappa_{12}>\kappa_\text{EP3}$, there are three distinct real eigenvalues. The above analysis demonstrates that the proposed system's eigenfrequency characteristics are consistent with those of a pseudo-Hermitian system \cite{mostafazadeh2002pseudo}. The third-order EP, \emph{i.e.}, $\kappa_{12} = \kappa_\text{EP3} = g \sqrt{(1+\alpha)/(2+\alpha)}$ is the point at which the three eigenvalues coalesce and become degenerate. Moreover, by tuning the scaling factor $\alpha$, the position of the EP3 can be adjusted to provide additional design flexibility. Even though we only consider the construction of EP3 in a 3-dimensional coupled resonator system, it should be noted that the proposed pseudo-Hermitian system can also be extended to higher-dimensional system where HOEPs are envisioned which has been theoretically validated in the supplementary material \cite{supplementary}.

Next, we analyze the perturbation behavior at the third-order EP of the proposed pseudo-Hermitian system, aiming to theoretically validate the enhanced sensing characteristics. When introducing a capacitive perturbation on all three resonators, the perturbation Hamiltonian $\hat{H}_p$ can be written as
\begin{equation} \label{eq:hp}
    \hat{H}_p = \frac{1}{2}\begin{pmatrix}
        \epsilon_1   &   0            &   0  \\
        0            &   \epsilon_2   &   0   \\
        0            &   0            &   \epsilon_3 
    \end{pmatrix}.   
\end{equation}
Here, we only consider the case when a capacitive perturbation on the relay resonator is introduced, \emph{i.e.}, $\epsilon_1 = \epsilon_3 = 0$ and $\epsilon_2 = \epsilon$, where the perturbation parameter $\epsilon = -\Delta C/(C+\Delta C)$ is related to the capacitive perturbation $\Delta C$. For the same gain parameter $g$, the frequency response is the most significant when perturbing the relay resonator. The other two scenarios are discussed and compared in the supplementary material \cite{supplementary}. By solving for the eigenvalues of the perturbed system with the Hamiltonian $\hat{H} = \hat{H}_0 + \hat{H}_p$, the perturbation behavior of the system at EP3 can be determined. The characteristic equation $\det(\hat{H}_0 + \hat{H}_p - \omega \mathbb{I}_3) = 0$ at EP3 when $\kappa_{13} = 0$ yields
\begin{equation} \label{eq:perturbed_char_eq}
    \lambda^3 - \frac{\epsilon}{2}\lambda ^2 + \text{i}\frac{\alpha g}{4(1+\alpha)}\epsilon\lambda -\frac{g^2}{8(1+\alpha)} \epsilon = 0.
\end{equation}
By expanding the eigenvalues with a Newton-Puiseux series \cite{hodaei2017enhanced} $\lambda = c_1 \epsilon^{1/3} + c_2 \epsilon^{2/3} + \ldots$ and substituting it into the characteristics equation \eqref{eq:perturbed_char_eq}, we have
\begin{equation} \label{eq:expand}
\begin{split}
    c_2^2 \epsilon^{\frac{7}{3}} + 2(c_1 c_2 - c_2^3) \epsilon^2 + 2({c_1^2}/{2}-3 c_1 c_2^2-\text{i}\beta c_2 g) \epsilon^{\frac{5}{3}} \\ 
    - 2(3 c_1^2 c_2+\text{i}\beta c_1 g) \epsilon^{\frac{4}{3}}  + (g^2 \beta/\alpha - 2 c_1^3) \epsilon = 0,
  \end{split}
\end{equation}
where $\beta = \alpha/[4(1+\alpha)]$. Solving the last two terms of \eqref{eq:expand}, \emph{i.e.}, $g^2 \beta/(2\alpha)-c_1^3 = 0$ and $3 c_1 c_2+\text{i}\beta g = 0$, the coefficients $(c_1,c_2)$ of the Newton-Puiseux series can be obtained, which are $c_1 = (1+\alpha)^{-1/3} g^{2/3}/2$ and $c_2 = -\text{i}\alpha (1+\alpha)^{-2/3} g^{1/3}/6$. It can be seen that both $c_1$ and $c_2$ are decreasing functions of the scaling factor $\alpha$ for a fixed $g$. Therefore, as $\alpha$ increases, the bifurcation of the perturbed eigenvalue will decrease, resulting in an reduction in sensitivity. Thus, in practical use, the loss parameter of the second resonator should be as small as possible. Details of the circumstances under which $\alpha$ is taken at different values are discussed in the supplementary material \cite{supplementary}. Here, we only consider the case when $\alpha = 1$, \emph{i.e.}, same loss for the two lossy resonators. 

\begin{figure}[ht!]
    \centering
    \includegraphics[width = 3.2in]{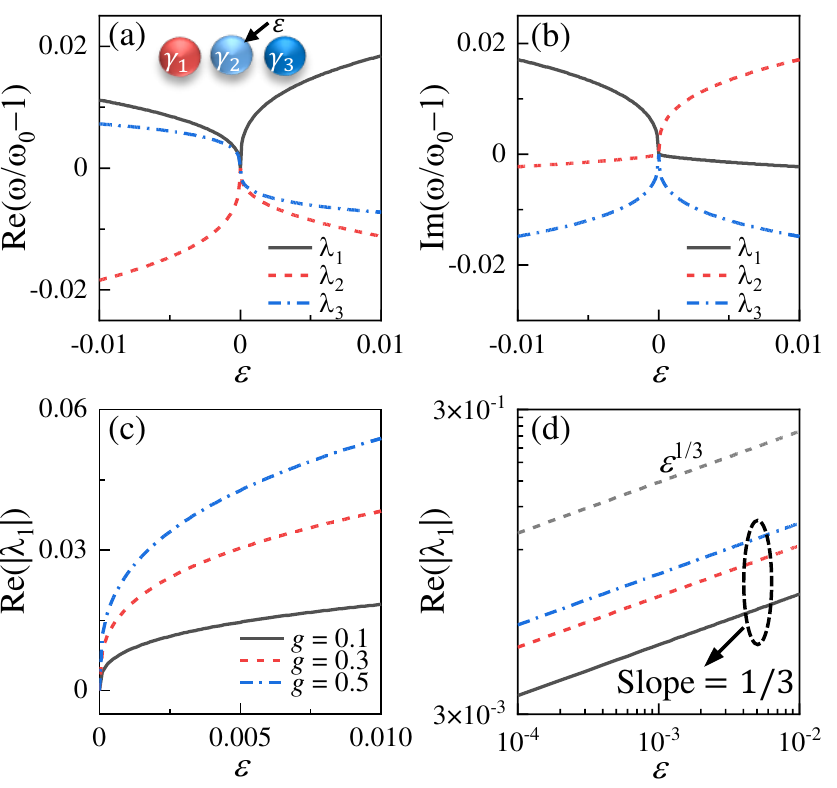}
    \caption{Theoretical result of the (a) real and (b) imaginary parts of the perturbed eigenfrequencies around EP3 when $g = 0.1$. (c) The bifurcation behavior of $\lambda_1$ at different $g$ when $\epsilon \in (0,0.01)$, and (d) the corresponding logarithmic behavior of each curve.}
    \label{fig:fig02}
\end{figure}

The characteristic equation at EP3 when $\alpha = 1$ yields $\lambda^3 - \epsilon\lambda^2/2 + \text{i}g\epsilon\lambda/8 - g^2\epsilon/16 = 0$, by solving which the following set of perturbed eigenfrequencies can be obtained, \emph{i.e.},
\begin{subequations} \label{eq:perturbed_eigenvalues}
\begin{align}
    \lambda_1 & = (g^2\epsilon/16)^{1/3} - \text{i} (g\epsilon^2/4)^{1/3} /6 + \ldots,  \\
    \lambda_2 & = (g^2\epsilon/16)^{1/3}e^{\text{i}2\pi/3} - \text{i} (g\epsilon^2/4)^{1/3} e^{-\text{i}2\pi/3}/6 + \ldots,  \\
    \lambda_3 & = (g^2\epsilon/16)^{1/3}e^{-\text{i}2\pi/3} - \text{i} (g\epsilon^2/4)^{1/3} e^{\text{i}2\pi/3}/6 + \ldots.
\end{align}
\end{subequations}
 The real and imaginary parts of the perturbed eigenvalues \eqref{eq:perturbed_eigenvalues} when $g = 0.1$ are plotted in \figref{fig:fig02}(a) and \figref{fig:fig02}(b), respectively.

As shown in \figref{fig:fig02}, the eigenfrequencies bifurcate from a single point at EP3 to three different complex numbers when the perturbation is applied. Moreover, the frequency response follows a cube-root dependence on the capacitive perturbation, which is consistent with the perturbation behavior of a third-order EP. It can be confirmed by the logarithmic behavior in \figref{fig:fig02}(d) which shows a straight black curve with a slope of $1/3$. Since the system exhibits three different modes, there will be three dips on the measured reflection spectrum curve, which will be validated in detail later. The $Q$-factor of a resonator circuit can be defined based on complex frequency, \emph{i.e.}, $Q = \text{Re}(\omega)/[2\text{Im}(\omega)]$ \cite{pozar2011microwave}, indicating that a smaller imaginary part of eigenfrequency results in a larger $Q$-factor and thus higher spectral resolution. Two of the three dips corresponding to the modes with a larger imaginary part will be hard to distinguish, so it will be challenging to accurately extract the dip frequencies. One of the three dips corresponding to the mode with the least imaginary part, \emph{i.e.}, the least gain or loss, will be narrow and sharp and can be used for the dip frequency measurement. As can be seen in \figref{fig:fig02}(b), the mode with the least imaginary part corresponds to the red dashed line $\lambda_2$ when $\epsilon < 0$ and the black solid line $\lambda_1$ when $\epsilon > 0$. Furthermore, the above-mentioned modes are only real when $\epsilon = 0$. Apart from that, the imaginary part has a small absolute value, indicating that the reflection spectrum is sharpest at EP3, but the sharpness will degrade as $\epsilon$ increases. We also compare the eigenfrequency response at different gain parameters in \figref{fig:fig02}(c), as can be seen from \eqref{eq:perturbed_eigenvalues} that the bifurcation is dependent on the gain parameter $g$. The logarithmic behavior of each case is illustrated in \figref{fig:fig02}(d), together with the gray dashed curve $\epsilon^{1/3}$ as a reference. It is shown that the cube-root dependence on the perturbation remains the same for different $g$, but the significance of the eigenfrequency bifurcation increases as $g$ increases. The value of the gain is related to the coupling condition, as has been analyzed. Specifically, the larger the gain value, the larger the coupling coefficients $\kappa_{12},\kappa_{23}$. On the one hand, during the experiment, stronger coupling will inevitably introduce coupling between nonadjacent resonators. On the other hand, strong coupling will make the theoretical result inaccurate since CMT is based on a weak coupling approximation. Therefore, in the theoretical analysis and subsequent experimental verification, the gain parameter is taken as 0.1 for a proper coupling condition.

\begin{figure}[ht!]
    \centering
    \includegraphics[width = 3.3in]{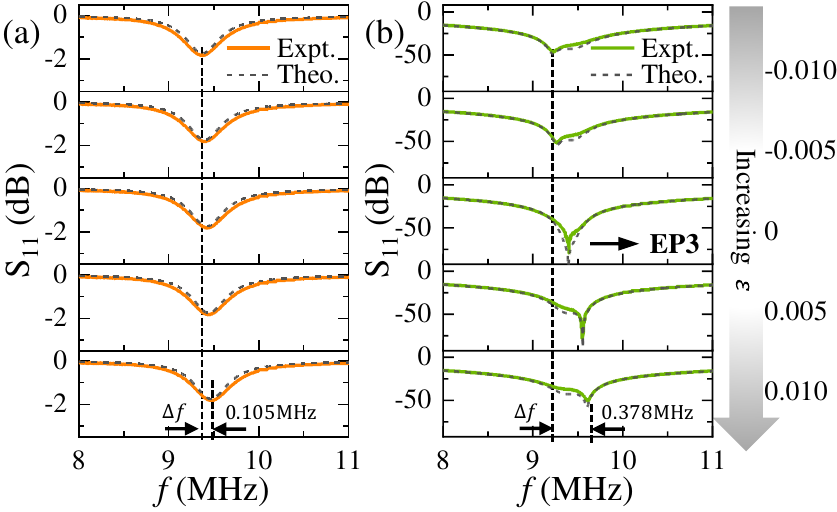}
    \caption{Experimentally measured reflection spectra (solid curve) of (a) conventional system when $\kappa = 0.1$ and (b) pseudo-Hermitian system around EP3 when $\kappa_{12} = 0.0816,\kappa_{23} =0.0289$ in comparison with the theoretical result (dashed curve) as the capacitive perturbation $\epsilon$ varies from $-0.01$ to $0.01$. The gain parameter is $g = 0.1$.}
    \label{fig:fig03}
\end{figure}

\begin{figure}[ht!]
    \centering
    \includegraphics[width = 3.0in]{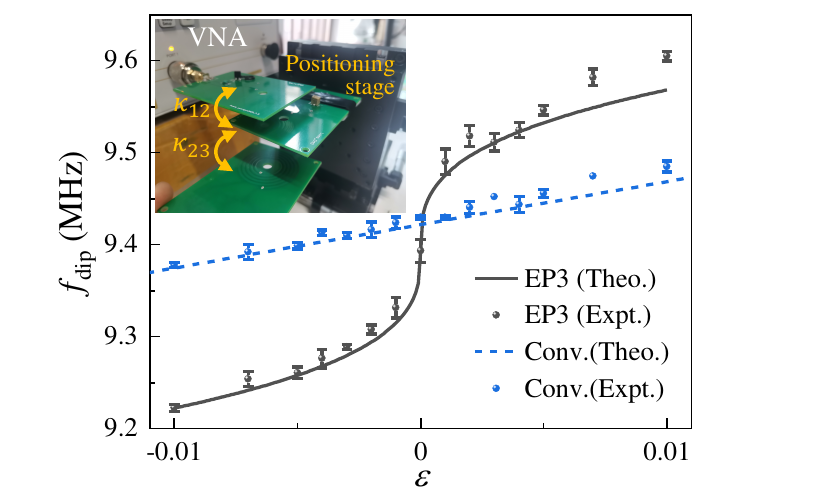}
    \caption{Comparison between theoretical (curves) and experimental (markers) results of conventional and EP3 sensing schemes. The inset shows the experimental setup of the proposed pseudo-Hermitian sensing system.}
    \label{fig:fig04}
\end{figure}

A printed circuit board (PCB)-based prototype of the pseudo-Hermitian sensing system was constructed for experimental validation. The experimental setup is depicted in the inset of \figref{fig:fig04}. Printed spiral planar inductors with inductances of $L=0.847~\si{\micro\henry}$ are manufactured on three PCBs that are coaxially aligned; and the adjacent vertical distances are adjusted with a positioning stage for coupling parameter tuning (for more information on the coil design, see Ref.~\onlinecite{yin2022ultrahigh}). Resonator I is connected to a KEYSIGHT E5063A VNA in order to measure the reflection spectrum $S_\text{11}$ . Since the VNA itself is a microwave source with an intrinsic impedance of $Z_0 = 50~\si{\ohm}$, it is analogous to a negative resistor whose value equals to $Z_0$. Therefore, Resonator I and the VNA constitute a gain resonator with negative resistance if $R_1$ is less than $50~\si{\ohm}$. In our experiment, the circuit parameters are as follows: $L = 0.847~\si{\micro\henry}$, $R_\text{2} = R_\text{3} = R = 2.5~\si{\ohm}$, $R_\text{1} = Z_0 - 2R = 45~\si{\ohm}$; and a variable capacitor is used to mimic a capacitive sensor. When $g = 0.1$ and $\alpha = 1$, the calculated coupling condition at EP3 is $\kappa_{12} = 0.0816$ and $\kappa_{23} = 0.0289$. The capacitance of $C_\text{EP3} = 338.8~\si{pF}$ at EP3 determines the preset capacitance of the three LC resonators. When the perturbation parameter changes from $-0.01$ to $0.01$, the relay resonator's capacitance changes from $342.222~\si{pF}$ to $335.446~\si{pF}$. The experiment results are shown in \figref{fig:fig03}.

Figures \ref{fig:fig03}(a) and \ref{fig:fig03}(b) compare the measured reflection spectra of the conventional and the proposed pseudo-Hermitian readout schemes as the capactive perturbation varies. The theoretical $S_{11}$ parameter is calculated using the formula $S_{11} = 20 \log_{10} \left|(Z_\text{in}-Z_0)/(Z_\text{in}+Z_0)\right|$, where $Z_\text{in}$ is the input impedance from the terminal of the VNA. The theoretical derivation for the reflection coefficient of both the conventional and the proposed systems is provided in the supplementary material \cite{supplementary}. Figure \ref{fig:fig03}(a) demonstrates that the dip of the reflection spectra for the conventional readout scheme is shallow, in which case the minimum of the $S_{11}$ curve exceeds $-2~\si{dB}$. In contrast, the reflection dip is much deeper for the proposed system and the minimum of $S_{11}$ can reach several tens of decibels. Consequently, the proposed system has greater spectral resolution than the conventional scheme. Furthermore, \figref{fig:fig03}(b) demonstrates that the resolution of the spectrum dip is greatest at EP3 but degrades as the perturbation increases due to the existence of an imaginary part when $\epsilon \neq 0$ [see \figref{fig:fig02}(b)]. As the perturbation increases, so does the value of the imaginary part, leading to a reduction in the $Q$-factor and thus a loss of spectral resolution. Moreover, the resolution is greater for positive ($\Delta C < 0$) than negative perturbation ($\Delta C > 0$), indicating an asymmetric $Q$-factor under symmetric perturbation. This asymmetries phenomenon originate primarily from the approximation in CMT. It should also be noted that the experimental $S_{11}$ spectrum curve at EP3 is not as smooth as the theoretical curve due to the superposition of a noise signal, which is primarily the thermal noise of the circuit \cite{xiao2019enhanced}.
Regarding the sensitivity, as shown in \figref{fig:fig04}, the eigenfrequency evolution in response to the perturbation parameter at EP3 follows a cube-root dependency on the capacitive perturbation. Compared to the conventional scheme, where a linear dependency is observed, the sensitivity of the proposed pseudo-Hermitian-EP3-based readout scheme is significantly enhanced. In the experiment, for the same sensor capacitance change $\Delta C = 6.776~\si{pF}$, the dip frequency change for the proposed system is $0.378~\si{MHz}$, while it is only $0.105~\si{MHz}$ for the conventional scheme. Moreover, the sensitivity enhancement is higher as $\epsilon$ approaches zero.

In summary, we proposed a more general pseudo-Hermitian system based on a radio-frequency circuit that can realize third-order EP under an asymmetric adjacent coupling condition and arbitrary loss parameters. A prototype is built to demonstrate the theoretical proposal experimentally. A cube-root dependency on the perturbation for the perturbed eigenfrequencies is observed when capacitive perturbation is introduced on the Resonator II (relay resonator), showing a more significant response than perturbing Resonators I and III. The sensitivity of the proposed system is much higher compared to the conventional wireless readout method, with the greatest sensitivity enhancement appearing at EP3. Compared with the conventional wireless sensing system, the proposed circuit shows enhanced resolution of the measured reflection spectrum. Our work demonstrates that HOEP-enhanced sensing can be utilized under arbitrary coupling and gain-loss parameters, given that the pseudo-Hermitian condition is satisfied, showing potential applications in scenarios such as wireless intraocular pressure monitoring, parameter detection in harsh environments, \emph{etc}.

\begin{acknowledgments}
T.D. is grateful to the National Natural Science Foundation of China (NSFC) for their funding under grant no. 51977165.
\end{acknowledgments}

\vspace{-1.0em}
\bibliography{main}

\end{document}